\documentclass[a4paper,fleqn]{cas-sc}

\usepackage[numbers]{natbib}
\DeclareUnicodeCharacter{2212}{-}
\usepackage[export]{adjustbox}
\usepackage{xcolor}
\usepackage{caption}
\def\tsc#1{\csdef{#1}{\textsc{\lowercase{#1}}\xspace}}
\tsc{WGM}
\tsc{QE}
\tsc{EP}
\tsc{PMS}
\tsc{BEC}
\tsc{DE}
\begin{document}
\let\WriteBookmarks\relax
\def\floatpagepagefraction{1}
\def\textpagefraction{.001}

\shorttitle{Empathy Models and Software Engineering}

\shortauthors{Hashini Gunatilake et~al.}

\title [mode = title]{Empathy Models and Software Engineering - A Preliminary Analysis and Taxonomy}                      
% \tnotemark[1,2]

\tnotetext[1]{Gunatilake, Grundy and Mueller are supported by ARC Laureate Fellowship FL190100035.}

% \tnotetext[2]{The second title footnote which is a longer text matter
%    to fill through the whole text width and overflow into
%    another line in the footnotes area of the first page.}

\author[1]{Hashini Gunatilake} [auid=000,bioid=1,orcid=0000-0002-4823-0214]

\cormark[1]


\ead{hashini.gunatilake@monash.edu}

\credit{Conceptualization, Writing - Original Draft Preparation, Review \& Editing}

\address[1]{Department of Software Systems and Cybersecurity, Faculty of Information Technology, Monash University, Melbourne, Australia}

\author[1]{John Grundy}
\credit{Conceptualization, Writing - Review \& Editing}

\author[1]{Ingo Mueller}
\credit{Conceptualization, Writing - Review \& Editing}

\author[1]{Rashina Hoda}
\credit{Conceptualization, Writing - Review \& Editing}

\cortext[cor1]{Corresponding author}
% \cortext[cor2]{Principal corresponding author}

% \fntext[fn1]{This is the first author footnote. but is common to third
% \fntext[fn2]{Another author footnote, this is a very long footnote and
%   it should be a really long footnote. But this footnote is not yet
%   sufficiently long enough to make two lines of footnote text.}

% \nonumnote{This note has no numbers. In this work we demonstrate $a_b$
%   the formation Y\_1 of a new type of polariton on the interface
%   between a cuprous oxide slab and a polystyrene micro-sphere placed
    
\begin{abstract}
Empathy is widely used in many disciplines such as philosophy, sociology, psychology, health care. Ability to empathise with software end-users seems to be a vital skill software developers should possess. This is because engineering successful software systems involves not only interacting effectively with users but also understanding their true needs. Empathy has the potential to address this situation. Empathy is a predominant human aspect that can be used to comprehend decisions, feelings, emotions and actions of users. However, to date empathy has been under-researched in software engineering (SE) context. In this position paper, we present our exploration of key empathy models from different disciplines and our analysis of their adequacy for application in SE. While there is no evidence for empathy models that are readily applicable to SE, we believe these models can be adapted and applied in SE context with the aim of assisting software engineers to increase their empathy for diverse end-user needs. We present a preliminary taxonomy of empathy by carefully considering the most popular empathy models from different disciplines. We encourage future research on empathy in SE as we believe it is an important human aspect that can significantly influence the relationship between developers and end-users.

\end{abstract}

\begin{highlights}
\item We developed a preliminary taxonomy of empathy considering widely used empathy models
\item We present our vision on empathy studies in software engineering(SE) context
\item We discuss suitability \& acceptability of different empathy models in SE context
\end{highlights}

% Keywords

\maketitle
\section{Introduction} 
\label{introductionSection}

Software is built for people by people. In the SE discipline, we try to make software more inclusive and user friendly by respecting the diversity of users. However, what makes this process difficult is the diverse nature of people. Understanding the needs and expectations of somebody can be difficult, especially if the other person is different to us, e.g. in terms of age, gender, culture, education, etc. However, it is essential that software developers take full cognisance of these differences in order to adequately meet diverse user needs. Empathy is one such core human aspect that may have a great influence on the SE process.

Engineering successful software systems involves not only interacting effectively with other stakeholders but also understanding their true needs. Empathy provides a way to help accomplish this challenging mission \citep{Pina2021}. Much future software technology is adopting AI-based software, and technology companies like Google and Grammarly have adopted empathy into their key products \citep{Allegretti2018, Fluckinger2021}. Google has emphasised the importance of empathy in technology by establishing Google Empathy Lab to train empathy into Google’s algorithms \citep{Johnson2018, Allegretti2018}. One of their major goals is to make Google Assistant more empathetic in order to build a closer connection with users, by trying to understand them better than a typical AI. Grammarly Inc. has introduced a feature to their Grammarly Business product to set the tone for written communications with the purpose of providing more empathetic responses to the customers \citep{Fluckinger2021}. These attempts of technology organisations to embed empathy into their products emphasise the vital role played by empathy in technology. %Hence we argue it is important to understand the role of empathy in software engineering context which has a close relationship with technology.   

Empathy can help a young developer to better understand how a much older person uses a smart home, who may have a completely different set of expectations and needs. Empathy can assist a developer with good eyesight to better understand how a visually-impaired person using a mobile application may have challenges with its use. Empathy is thus useful for software developers in understanding how diverse the users can be and the heterogeneous needs they may have. Empathy may also help a developer to act more effectively with designers when there are limitations in the technology to implement a proposed design. If developers are sufficiently empathetic, then they may be able to discuss the infeasible aspects in the design and find alternative technical solutions for these aspects without rejecting the entire design, which might have taken weeks for the designers to complete. Likewise practicing empathy can help the developers to eliminate team conflicts and work harmoniously to identify the best solutions which would ultimately lead to better software \citep{lundstrom2015perceptions}. Similarly there are many instances a software developer can apply empathy throughout different stages of software development which include empathy for end-users (this being our central focus), teammates, future developers, and other stakeholders \citep{Pina2021}. Each of these instances are further discussed below to emphasise the usefulness and significance of empathy in SE context.   

End-users of the same software application can be widely dissimilar and their needs can be divergent; some users might be colour-blind, have problems with the vision, have accessibility issues, have technological challenges, be too old or too young to understand the software functionality. Some might even feel insecure in using the software because of data breaches and privacy concerns \citep{Pina2021}. We define “end-users” as the people who utilise computer systems for their day-to-day tasks, occupational purposes and entertainment, as well as for overcoming disabilities or other challenges \citep{costabile2007visual}. The produced software might not serve the needs of end users if the software developer fails to sufficiently understand and consider their differences. 
Software developers need the ability to empathise with their end-users during different phases of the software development life cycle (SDLC). 

Empathy for teammates is another instance where developers can apply empathy. Teamwork and empathy are known to have a very close relationship \citep{Pina2021, akgun2015antecedents}. Software developers closely work with requirement engineers, designers, other developers, testers, etc. Developers may have to empathise with these different stakeholder groups primarily during planning, requirements elicitation, designing, development and testing phases. Empathising with different conditions of team members i.e., strengths, weaknesses, physical conditions, mental conditions, is crucial in order to maintain good team dynamics, which ultimately leads to greater performance and better software \citep{Pina2021,lundstrom2015perceptions, akgun2015antecedents}. 

Empathy for future developers is also seem important in SE context \citep{Pina2021}. Sometimes novice software developers spend an extensive amount of time trying to understand the code written by a previous developer. Developers can be empathetic towards future developers by increasing readability of the code through inline comments and following the coding best practices. This will not just save time but also save resources in software development project. Empathy for other stakeholders, such as the leadership team and other teams within the organisation, is important for useful flow of information and collaboration \citep{Pina2021}. In an organisation, there may be different teams who depend on the output of a particular team. In such cases, if this specific team can empathise with the needs and challenges of other teams, then this team may put the best possible effort to complete the intended tasks within the planned timeline. This may make the life easier for other teams and it may lead to good inter-team relationship. All these forms of applying empathy by a software developer will make the SE context better by assisting in addressing the diverse needs of users, improving the quality of relationships, saving time and even preserving resources. Hence, empathy provides a competitive advantage to SE context and it assists in building better software by addressing the real needs of the users \citep{Pina2021}. Despite the growing interest in human aspects and their impact on SE, research into the use of empathy in SE is still in its early stages. Only a few research studies have been conducted on the use of empathy in SE.

In this paper, we present a preliminary review and analysis of key empathy-related work and models, with a view to addressing this under-researched aspect in SE research. We then offer a preliminary taxonomy of empathy based on an analysis of the available conceptual and pragmatic models of empathy, techniques, and measures of empathy, to help guide future researchers. We first survey related work of empathy in SE and other domains (Section \ref{relatedWorkSection}) then we present our preliminary taxonomy of empathy (Section \ref{taxonomySection}). Next we summarise some key models (Section \ref{modelsOfEmpathySection}) and major techniques of empathy (Section \ref{empathyTechniquesSection}) from diverse disciplines which can be possibly adopted to the field of SE. Then we explore several leading measures of empathy used in different disciplines (Section \ref{empathyMeasuresSection}). Based on our analysis of empathy, we discuss different aspects for empathy research in SE (Section \ref{discussionSection}). Then we suggest directions for future research in empathy in the field of SE (Section \ref{ourVisionSection}). This will lastly be followed with a summary of our research (Section \ref{summarySection}).

\section{Related Work} 
\label{relatedWorkSection}

Empathy has been used to varying degrees in different fields. In medicine and nursing, empathy is heavily used and considered as one of the vital elements such professionals should cultivate \citep{hojat2018jefferson} \citep{yu2009evaluation}. In contrast, empathy is a relatively novel approach for some fields like professional writing and reviewing \citep{de2007professional}. We review some key studies from related literature to demonstrate the breadth and depth of applying empathy. 

\subsection{Complex Nature of Empathy}
\label{empathyRelatedWorkSection}
Empathy is a multi-dimensional construct with an extensive number of definitions which makes it challenging for researchers to reach a consensus on an unified definition \citep{clark2019feel, cuff2016empathy, neumann2015measures}. Even though it is not easy to choose one specific definition, we consider empathy as “\textit{understanding a person from his or her frame of reference rather than one’s own, or vicariously experiencing that person’s feelings, perceptions, and thoughts}” \citep{apa}. This is because, this definition best reflects the need for empathy in a software engineering context as per our understanding.
Different humans experience and respond to empathy in divergent ways. Many studies have argued that empathy operates in multiple levels as a trait as well as a state \citep{clark2019feel, cuff2016empathy}. In the domain of psychology, \emph{traits} are considered as relatively stable and enduring characteristics \citep{nezlek2007naturally} hence trait empathy is apprehended as a person’s stable character trait which is not influenced by external factors \citep{sep-empathy}. On the other hand, \emph{states} are supposed to change with time or with response to situational cues. Therefore state empathy is understood as outcomes or reactions in a specific situation \citep{nezlek2007naturally, sep-empathy}. Researchers have further stated that empathy varies within individuals over time, even within the same day \citep{clark2019feel}. Hence in some instances, the same person might empathise differently with the same situation, which makes it a complex construct to study. 

Apart from the multi-level components of empathy, many studies have acknowledged the importance of associating the multidimensional conceptualizations of empathy in research, otherwise the findings might not reflect empathy as a whole and potentially useful information might be excluded  \citep{clark2019feel, cuff2016empathy}. Some studies have argued that there are 3 dimensions to empathy -- Cognitive, Affective and Behavioural empathy \citep{clark2019feel}. Others have claimed there are only 2 dimensions as Cognitive and Affective empathy \citep{cuff2016empathy}. \emph{Cognitive empathy} is defined as “\textit{the tendency to understand, or the state of understanding, others' internal states}” \citep{clark2019feel}, which is often referred to as putting yourself in another person’s shoes. \emph{Affective empathy} is described as “\textit{feeling the same affective state as another person}”, which often refers to the unconscious tendency to share emotions of others \citep{wallmark2018neurophysiological}. \emph{Behavioural empathy} consists of two types of empathic behaviours i.e., behaviour mirroring and empathic communication \citep{clark2019feel}. Mimicking of facial expressions, mannerisms, postures, and gestures of the other person is referred as behaviour mirroring. Empathic communication is defined as intentional behaviour that displays cognitive and/or affective empathy towards the other person. State and trait multi-level phenomena and multi-dimensional manifestations are salient components in recent empathy research yet they make the empathy research more complex. This complicated nature of empathy research makes it even more complex to study empathy in SE. 

\subsection{Usage of Empathy in Software Engineering}
Although human aspects and impact on SE has been increasingly of interest in SE, studying the use of empathy in SE is still an emerging area. Only a very limited number of research studies have been conducted with regard to the use of empathy in SE. 

User experience (UX) is one such area where some complementary research on empathy has been conducted. Personas is a widely used design method which allows better understanding of users by describing the users’ characteristics, goals and skills \citep{ferreira2015designing}. In recent research studies \citep{ferreira2015designing} empathy map (EM) has been adopted for crafting personas in order to build a degree of empathy with the user. This improved the ability of the development team to understand the users and their real needs. During this study experiment all the participants agreed that the EM helped to better describe the personas. PATHY (Personas EmpATHY) is a technique which unites personas and EM and it not only describes the characteristics of personas but also provides an overview of the features that the software should have \citep{ferreira2016pathy}. Hence this technique guides software developers in better focusing on the application features based on personas. Design thinking (DT) is another frequently used problem solving approach in UX to develop solutions with a human centred approach. \citet{canedo2020design} has expressed the perception of software developers on use of DT in Agile software projects. In this study EM is presented as a tool used for the DT approach and their survey results showed that 15.4\% of the practitioners use EM as a DT tool in software development. 

However, another study highlights that usage of the EM method relies on the skills and experience of the facilitation experts \citep{bittner2019designing}. In this study, they tried to automate this process by developing a chatbot to act as the facilitation expert. During the evaluation of the developed chatbot, they have identified that the knowledge provided by chatbot on DT and the EM method has shown to be sufficient for conducting the EM method session. The researchers went to some length in trying to automate the DT process using the EM method. We might thus argue that the software provides a higher quality user experience when empathy is embedded in the design techniques, by better understanding the users.

Requirements elicitation (RE) is another noticeable space where empathy has been used. A study has stated that DT practice is a contributing factor to the increased quality of the software product delivered to the end user as this approach helps to understand the requirements more clearly prior to implementation \citep{canedo2020design}. Their survey results showed that 23.1\% of the practitioners use EM as a DT tool in requirements elicitation. Hence we can claim that empathy is a major contributor to the RE process. 

Levy and Hadar \citep{levy2018importance} explained how the empathy step in DT can be utilised by the developers to address privacy concerns of the users. In this study they used the Model of Empathy in engineering to highlight the instances where developers failed to practice empathy, also they have described how the perceptions of the software developers could have changed using the empathy step of DT which would lead to better understanding of requirements. These researchers have mostly used the EM when employing the DT approach yet they have also elaborated the importance of using personas and customer journey maps. They have stated that empathy capabilities of developers towards the end-users lead to a better defined set of specified and addressed privacy requirements which are critical in designing privacy-sensitive systems.

\subsection{Usage of Empathy Beyond Software Engineering}
Empathy research is far more common in non-software engineering fields like philosophy, sociology, psychology, medicine and nursing. Empathy is considered as one of the major elements in professionalism in medicine \citep{hojat2018jefferson}. Professional medical organisations have even endorsed cultivating empathy as one of the key goals in medical education. For instance, the enrichment of empathy is named as one of the educational objectives of medical schools by the Association of American Medical Colleges in the the Medical Schools Objectives Project \citep{aamc}. Further, many instruments to measure empathy have been developed in the context of health professions education and patient care \citep{hojat2018jefferson, shariat2013empathy}. Empathy is also regarded as an indispensable element of the nurse-patient relationship and is essential to quality nursing care \citep{yu2009evaluation}. Researchers have  developed numerous measures to assess empathy in nursing research \citep{yu2009evaluation}. In addition, there are many case studies where empathy has been successfully applied for education as a tool for improving abilities and competencies of Informatics Engineering students and Design students \citep{blanco2017deconstructing, levy2018educating}. Further application of empathy can also be seen in the domain of professional writing and reviewing \citep{de2007professional}. The researchers have summarised the shortcomings of not being able to empathise in professional writing and reviewing processes. Marketing is another field where empathy has been used to improve the performance and success of salespersons \citep{delpechitre2013review}. They also discussed the significance of a salesperson's empathy during a salesperson-customer interaction. Likewise empathy has been used as a complementary element and it is considered as an essential skill in diverse fields of research. 

{\subsection{Empathy Training}
Empathy training is an important area of research which has received more attention in fields like psychology, healthcare, and education. We did not find SE research studies related to empathy training. However inspired from the various studies discussed in this section, we see it as a potential avenue in improving empathy between developers and users. Having an understanding on the nature and the results of empathy training is significant in this study as empathy training is closely related to our vision on empathy in SE context. Thus, in this section we outline some of the empathy training studies in different contexts and factors which affect the efficacy of empathy training.}

{Researchers have examined empathy training to identify whether empathy can be taught. Some researchers have suggested empathy as a quality which is influenced by genetics \citep{butters2010meta, warrier2018genome}. However many studies were conducted to investigate the efficacy of empathy training in diverse settings and majority of the studies reported empathy training to be an effective process in promoting prosocial behavior \citep{teding2016efficacy, butters2010meta} which is defined as a form of positive psychology that exhibits behavior which is helpful to other people or society as a whole \citep{apa}. More emphasis has been placed on increasing the empathy levels of healthcare professionals particularly nurses, therapists, physicians and medical students \citep{dexter2012research}. A meta analysis on empathy training conducted by Dexter et al. discovered that empathy training had an overall positive effect. In a review conducted on 17 empathy training programs for nurses identified significant improvements in empathy in majority of the studies \citep{brunero2010review}. Majority of the studies reported considerable improvements in empathy levels of undergraduate medical students in a systematic review conducted on 18 studies \citep{batt2013teaching}. Another meta analysis on empathy training of medical students identified a positive effect in majority of empathy training programs \citep{stepien2006educating}. Several other meta analyses on psychotherapists' and helping professionals' empathy training reported moderate to large positive impact \citep{baker1989integrating, baker1990systematic}. Researchers have conducted many studies on empathy training in addition to those related to healthcare domain. A study focused on reducing prejudice among children via empathy training identified enhancement of their empathy levels upon receiving empathy training \citep{beelmann2014preventing}. Empathy training has also become vital in abuse prevention programs. Wiehe conducted a narrative review of empathy training programs to support in creating better child abuse prevention programs \citep{wiehe1997approaching}. This study reported that non-abusive parents displayed higher empathy compared to abusive parents. Further this study proposed a empathy training framework for parents and also recommended empathy training as a part of all parenting programs. The studies conducted on empathy training for rape prevention programs in college campuses discovered a reduction in future sexual assaults with the improvement of empathy \citep{foubert2006effects, o2003rape, lee1987rape}.}

{A meta analysis conducted on the efficacy of empathy training identified several variables which affects moderating empathy training \citep{teding2016efficacy}. This study has explored the type of trainees involved and it was identified that studies which involved health professionals and university students displayed significantly higher impact compared to youths or other types of adults. This finding is also supported by the outcomes of developmental and neurological research studies which discovered a particular degree of neurological maturity is required in order to properly comprehend and display empathy \citep{choudhury2006social, decety2006human}. This meta analysis has also explored the types of empathy trained and it was identified that all the included studies have at least considered cognitive empathy \citep{teding2016efficacy}. Researchers have argued this might be due to the nature of cognitive empathy to be involved in consciously acquired processes. Also during this meta analysis it was determined that studies with cognitive, affective, and behavioral, or cognitive and behavioral empathy displayed slightly higher effects compared to the studies with cognitive and affective empathy. Researchers have also analysed the objectivity of the outcome measures during this meta analysis. Significantly higher effects were observed in the studies with objective measures compared to those with self-report measures. These objective measures included written tests of ability to identify another person’s emotions and ratings of empathic behavior by patients. The findings of the meta analysis reported that empathy training using objective measures might lead to significantly higher positive impact compared to self-report measures. Further this meta analysis suggests more future research directions on empathy training to analyse the extent of enduring training benefits and the best empathy types combination which produces the most effective empathy training results. The overall results proved that empathy training programs are effective in increasing empathy levels which can be utilised to promote prosocial behavior \citep{teding2016efficacy}.}

\begin{figure}[h]
\centering
  \captionsetup{justification=centering,margin=3cm}
  \includegraphics[clip,scale=.45]{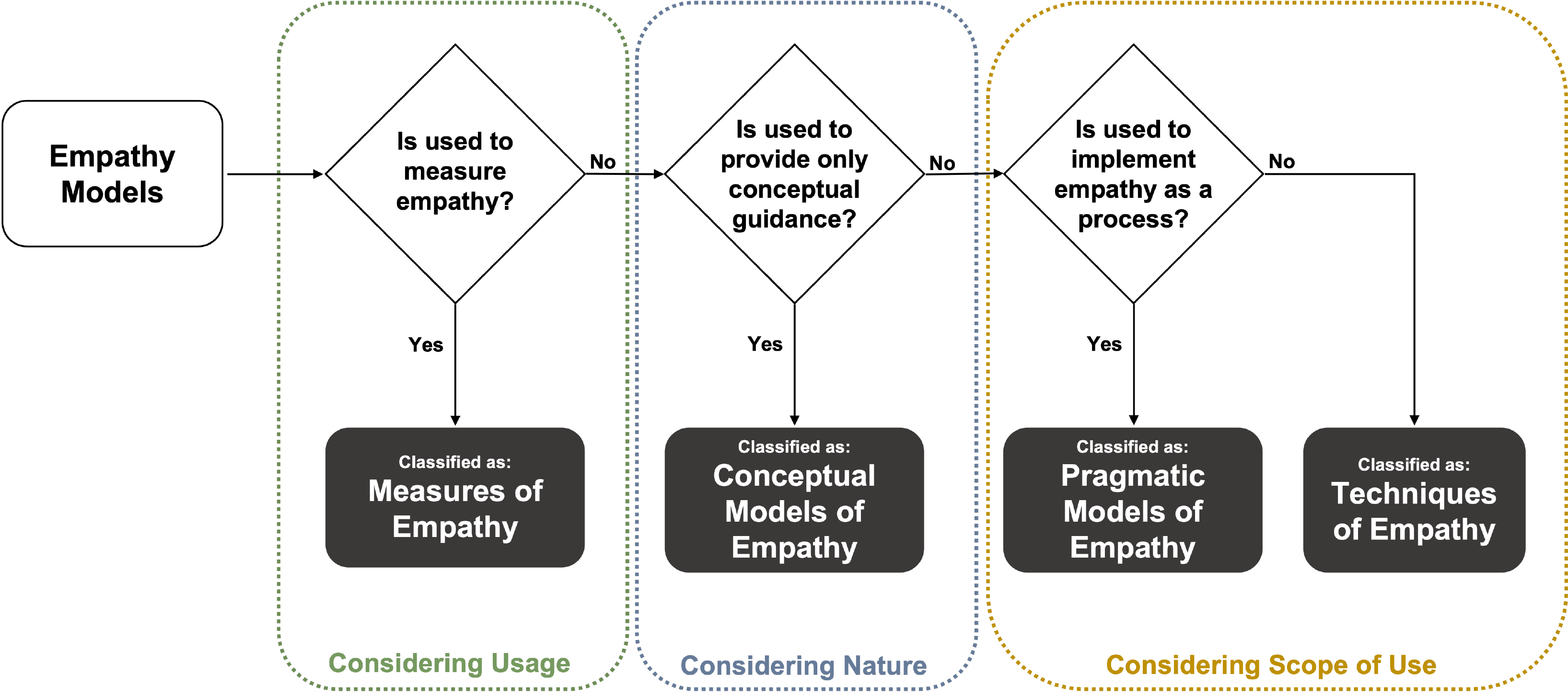}
  \caption{Development of a Preliminary Taxonomy of Empathy} 
  \label{classification}
\end{figure}

\begin{figure}[h]
\centering
  \includegraphics[scale=.51]{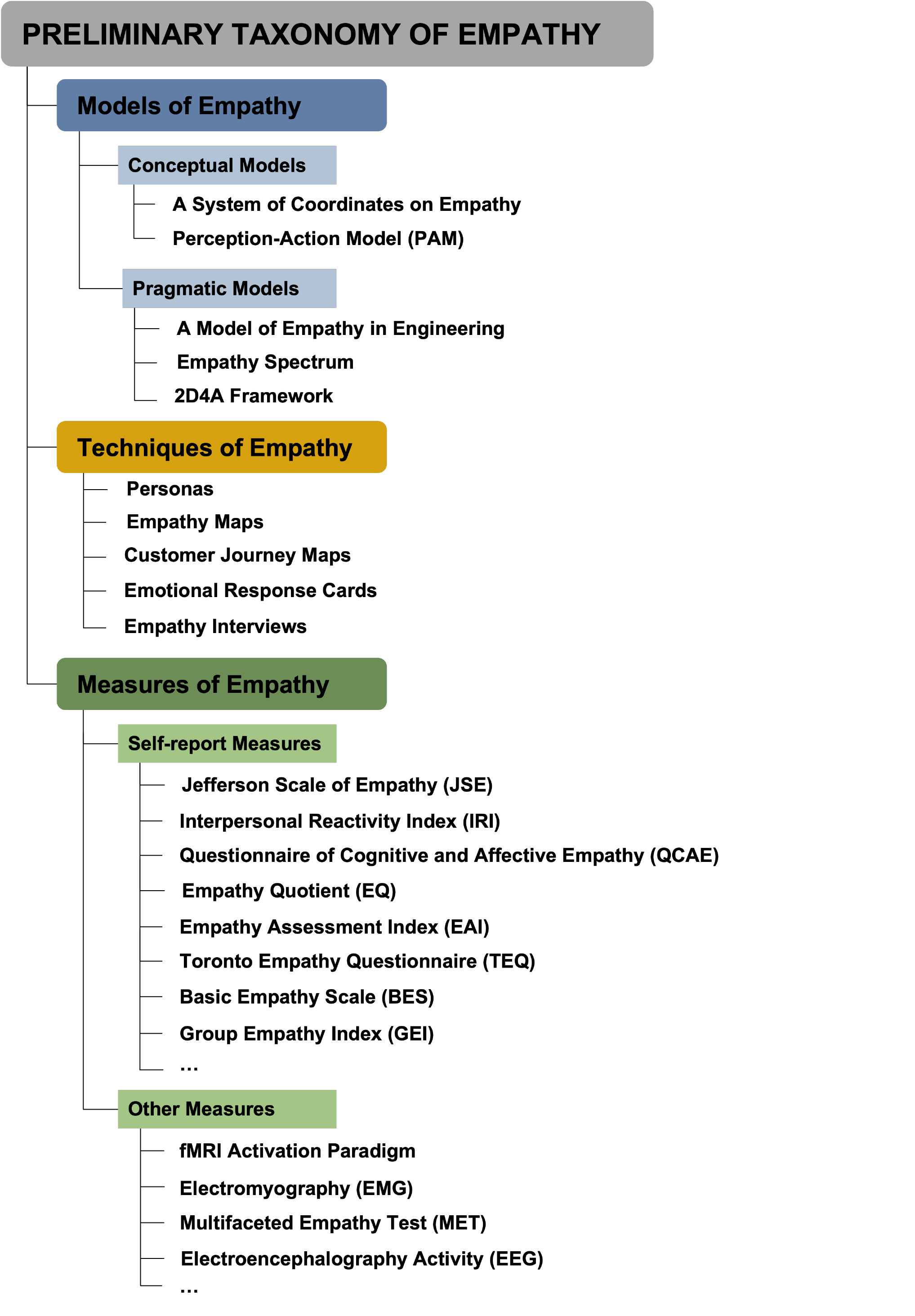}
  \caption{Preliminary Taxonomy of Empathy}
  \captionsetup{justification=centering}
  \label{taxonomy}
\end{figure}

\section{A Preliminary Taxonomy of Empathy} 
\label{taxonomySection}

{As we have seen in section \ref{empathyRelatedWorkSection} empathy refers to “\textit{understanding a person from his or her frame of reference rather than one’s own, or vicariously experiencing that person’s feelings, perceptions, and thoughts}” \citep{apa}. This section provides an overview of our taxonomy of empathy in SE. All the individual categories in the proposed taxonomy are described in detail in the following sections; Section \ref{modelsOfEmpathySection} explains the conceptual and pragmatic models of empathy, section \ref{empathyTechniquesSection} discusses the techniques of empathy and section \ref{empathyMeasuresSection} outlines the measures of empathy.}

We identified empathy as an under-researched concept to date in SE research due to the lack of studies focusing on empathy in an SE context. From our literature analysis, it was evident that a large number of diverse empathy models, frameworks, techniques and measures were produced to address different areas of empathy. However, we realised that, there was no proper taxonomy to distinguish among, and make sense, of these various models. {The lack of a proper taxonomy made it difficult to distinguish the available empathy models and understand how these empathy models can be applied in an SE context. We tried to address this gap by developing our taxonomy. Our goal of developing this taxonomy is to understand if and how the various empathy models, frameworks, techniques and measures can be used in the SE context. Also having a proper taxonomy is a preferable preliminary step to continue our research on empathy in SE. The final goal of developing this taxonomy is to make our knowledge accessible to SE researchers as empathy is an under-researched concept in the SE research community.}

We developed a preliminary taxonomy of empathy by analysing the various models available in the empathy literature and grouping them based on their usage, nature, and scope of use as illustrated in Figure \ref{classification}. {During our analysis, we identified four disjoint categories among these diverse empathy models. We determined that these various empathy models can be classified as the ones which are used to measure empathy and the ones which are not. Discovering whether there is an empathy scale that can be used to measure empathy in SE is one of the major goals we had while developing our taxonomy. Hence, our first categorisation was based on the “usage” of the empathy models. We classified the models which used to measure empathy as \textbf{Measures of Empathy}}.

{We then considered the remaining models which were not categorised as the measures of empathy. Among them, there were some models which assisted in better understanding empathy only as a concept and some other models which helped in implementing empathy in a practical setting. Hence we used the “nature” of the guidance provided by the empathy models as the next theme of our categorisation. We categorised the models provided only a conceptual guidance to empathy as \textbf{Conceptual Models of Empathy} which can also be introduced as descriptive models.}

{We noticed that the models which are not categorised as conceptual models of empathy are basically used to guide empathy building in practical settings. However, we noticed one difference among them i.e., the scope of applying these models in real life. Some of these models were designed to apply empathy as a complete process whereas some can only be used to apply empathy in an activity/phase/few different phases. Therefore, we used “scope of use” to further classify these models. We categorised the models which assisted in implementing empathy as a process as \textbf{Pragmatic Models of Empathy} which can also be identified as prescriptive models. The models which facilitated applying empathy only in an activity, a phase (a group of activities) or in a few different phases of a process were categorised as \textbf{Techniques of Empathy}.}

% We categorised the empathy scales as the measures of empathy by considering the usage aspect. Then, we considered the nature of the other available models and we determined that some can be used to provide practical guidance. Also, there were some models which can only be used as means of understanding empathy as a concept, hence, we classified them as conceptual models of empathy. Finally, we considered context \& scope of use aspect to check whether it is possible to classify the remaining models any further. We discerned that the models which provided practical guidance can be categorised further by considering the context \& scope of use. Empathy models which can be utilised to implement empathy as a process are categorised as pragmatic models of empathy, and the models which facilitated applying empathy in an activity, a phase (a group of activities) or in a few different phases of a process are categorised as techniques of empathy.

Our preliminary taxonomy of empathy contains key empathy models, techniques, and measures. The developed taxonomy which is illustrated in Figure \ref{taxonomy} is useful to get an understanding on the holistic view of empathy. {We identified two key conceptual empathy models during our analysis, namely \emph{A System of Coordinates on Empathy model} and \emph{Perception-Action Model}. Also we categorised \emph{Empathy Spectrum, A Model of Empathy in Engineering} and \emph{2D4A Framework}, as pragmatic models of empathy. We found several techniques that are used to apply empathy. \emph{Personas, Empathy Maps, Customer Journey Maps, Emotional Response Cards} and \emph{Empathy Interviews} are some of these key techniques. Further we found out many empathy scales that are used to measure empathy in different contexts. Some of the key empathy measures include the \emph{Jefferson Scale of Empathy, Interpersonal Reactivity Index, Questionnaire of Cognitive and Affective Empathy, Empathy Quotient, Empathy Assessment Index, Toronto Empathy Questionnaire, Basic Empathy Scale, Group Empathy Index, fMRI Activation Paradigm, Electroencephalogram, Multifaceted Empathy Test} and \emph{Electromyography}.} We have used the knowledge gained by studying existing literature to develop this taxonomy. It is a preliminary taxonomy which requires further refinement, especially for investigating applicability of empathy concepts to software engineering. Each of the categories in the taxonomy are described in detail in section \ref{modelsOfEmpathySection}, \ref{empathyTechniquesSection} and \ref{empathyMeasuresSection}.

\section{Models of Empathy} 
\label{modelsOfEmpathySection}
We found a broad range of empathy models which have been developed in different domains while analysing the literature and we picked some key models for the taxonomy. We identified that these models can be divided into two main categories i.e., conceptual models and pragmatic models.

\subsection{Conceptual Models of Empathy}

We categorised the empathy models which only influence the thought process of a person as conceptual models of empathy. These models are difficult to use as frameworks in real life use cases, which require guidance to make the current thinking state better by incorporating empathy. Rather these models are more like concepts that can be adopted for deeper understanding about the subtle nature of empathy. We identified some key conceptual empathy models used in different domains of research.

\begin{itemize}
    \item \textit{A system of coordinates on empathy}
\end{itemize}

 A system of coordinates on empathy model (Fig 1, page 6, of \citep{dong2017empathy}) is developed as a reference for designers and researchers to comprehend the meaning of empathy in design \citep{dong2017empathy}. The concept of empathy has evolved in many aspects since its origin and the recent advancements in neuroscience provided references for understanding empathy \citep{dong2017empathy}. However they claim that there are still  ambiguities about empathy and that researchers have so far failed to come to an agreement on the nature of empathy. They have developed a system of coordinates of empathy by utilising three aspects of ambiguities -- affection and cognition, subject-oriented and object-oriented, and attitude and ability. There are three axes which indicate six possibilities: two possible primary views of empathy i.e., cognitive and affective components; the focus of empathy can be either an attitude or a technique where the attention is drawn to how the designing is done rather than why it is done; and the chosen role of the empathy which can be either giving priority to the experience of the subject or the object. The model does not judge the value of one side of the axes over the other but acts as a reference for researchers to reflect on their mindset when doing empathy research.

\begin{itemize}
    \item \textit{Perception-Action Model (PAM)}
\end{itemize}

\begin{figure}[h]
\centering
  \includegraphics[scale=.45]{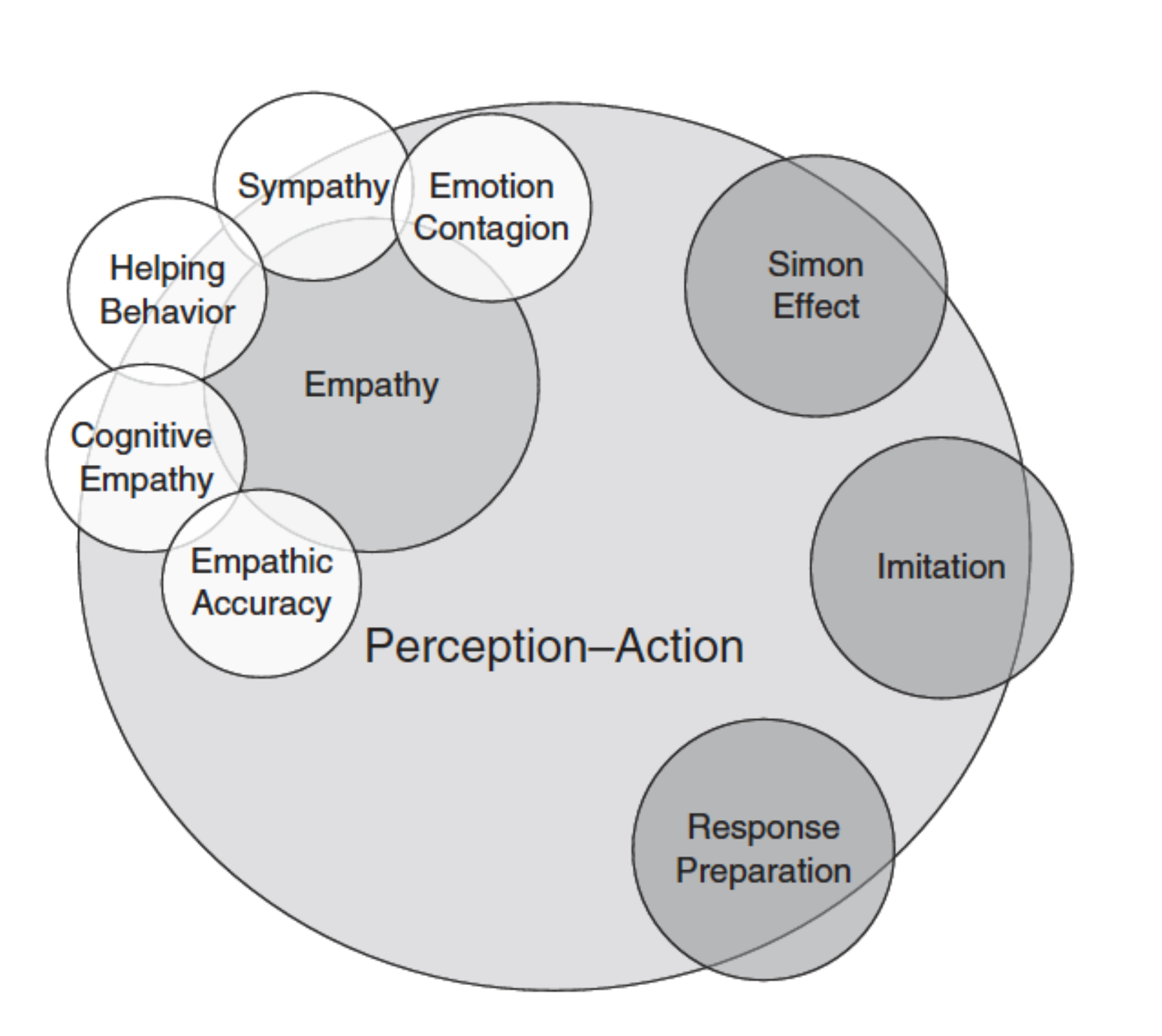}
  \caption{Perception-Action Model (from \citep{preston2012many})} 
  \captionsetup{justification=centering}
  \label{PAM}
\end{figure}

PAM (see Figure \ref{PAM}) of empathy states that empathy is a shared emotional experience which occurs when one person (the subject) starts to feel a similar emotion to another (the object) as a result of perceiving the other’s state \citep{preston2002empathy, preston2007perception, preston2012many}. It implies that the subject’s representations of the emotional state are automatically activated, upon the subject paying attention to the emotional state of the object. PAM is an attempt to unify various perspectives of empathy. There are two basic  levels of empathy included in PAM, namely motor behaviour and emotional behaviour. A perception-action mechanism also acts as the superclass of these categories. There are also several further sub-categories for these basic levels.

\subsection{Pragmatic Models of Empathy}
We categorised the models that can be used as frameworks to implement or incorporate empathy as a process, under pragmatic models of empathy. As the name implies these models provide more practical and realistic guidance. Pragmatic models help to understand how to integrate empathy as a process and how to put empathy into work. These models can be employed to implement empathy as a new process or integrate empathy into an existing process. During our analysis we analysed the following key frameworks which fall under this category.

\begin{itemize}
    \item \textit{A Model of Empathy in Engineering}
\end{itemize}

\begin{figure}[h]
\centering
  \includegraphics[scale=.5]{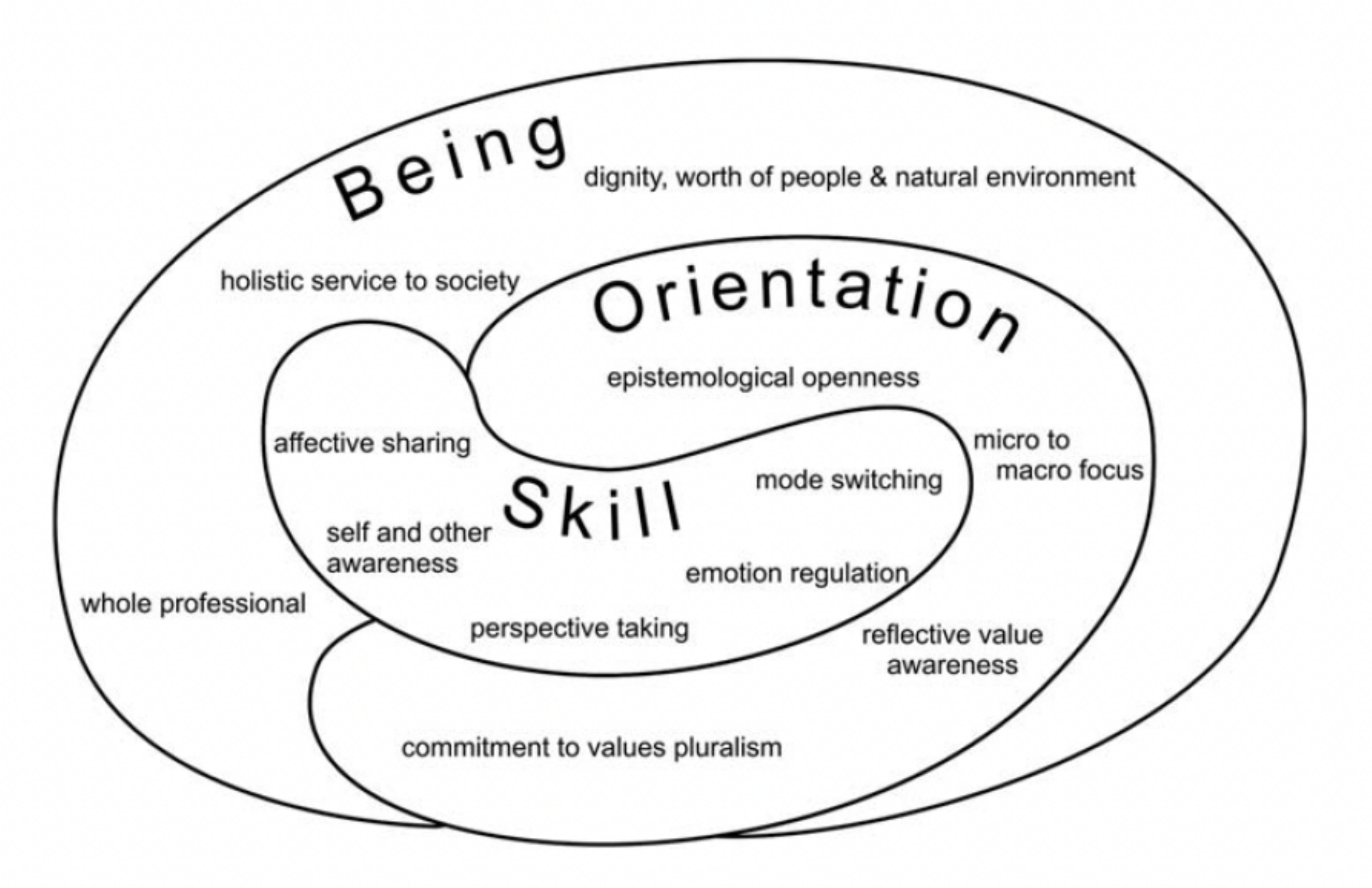}
  \caption{A Model of Empathy in Engineering (from \citep{walther2017model})} 
  \captionsetup{justification=centering}
  \label{engineeringmodel}
\end{figure}

This empathy model (see Figure \ref{engineeringmodel}) was developed especially for engineering and engineering education \citep{walther2017model}, but is also used heavily in other domains. This model conceptualises empathy in engineering as a teachable and learnable skill, a practice orientation, and a professional way of being. Researchers have demonstrated mutually dependent and supportive nature of each of these dimensions without using a hierarchical approach. In summary, the skill dimension consists of five unique socio-cognitive processes that interact with each other to support relationship building, empathic communication and decision making. The orientation dimension includes a set of mental characteristics that influence engineers’ engagement in practice situations. Lastly, the being dimension displays the need to establish practice orientations, empathic skills, and their development within a contextualising framework of broader values.

\begin{itemize}
    \item \textit{Empathy Spectrum}
\end{itemize}

Researchers have developed this holistic approach for empathy for nursing practice by considering empathy as a multifaceted phenomenon (Fig 1, page 3, of \citep{wiseman2007toward}). The model acknowledges four distinct forms of empathy including empathy as an incident, empathy as a way of knowing, empathy as a process, and empathy as a way of being. Empathy spectrum illustrates the different stages of empathy development and expression along the continuum, with empathy as an incident at one end and empathy as a way of being at the other end. The double helix along the continuum represents socialisation and knowledge, as socialisation and knowledge of nurses helps developing their empathy skills. The area around the continuum is referred as the context of care representing the role played by the context in the development and practice of empathy. It is possible to move in any direction along the continuum even though it is displayed as a linear diagram.

\begin{itemize}
    \item \textit{2D4A Framework}
\end{itemize}

\begin{figure}[h]
\centering
  \includegraphics[scale=.45]{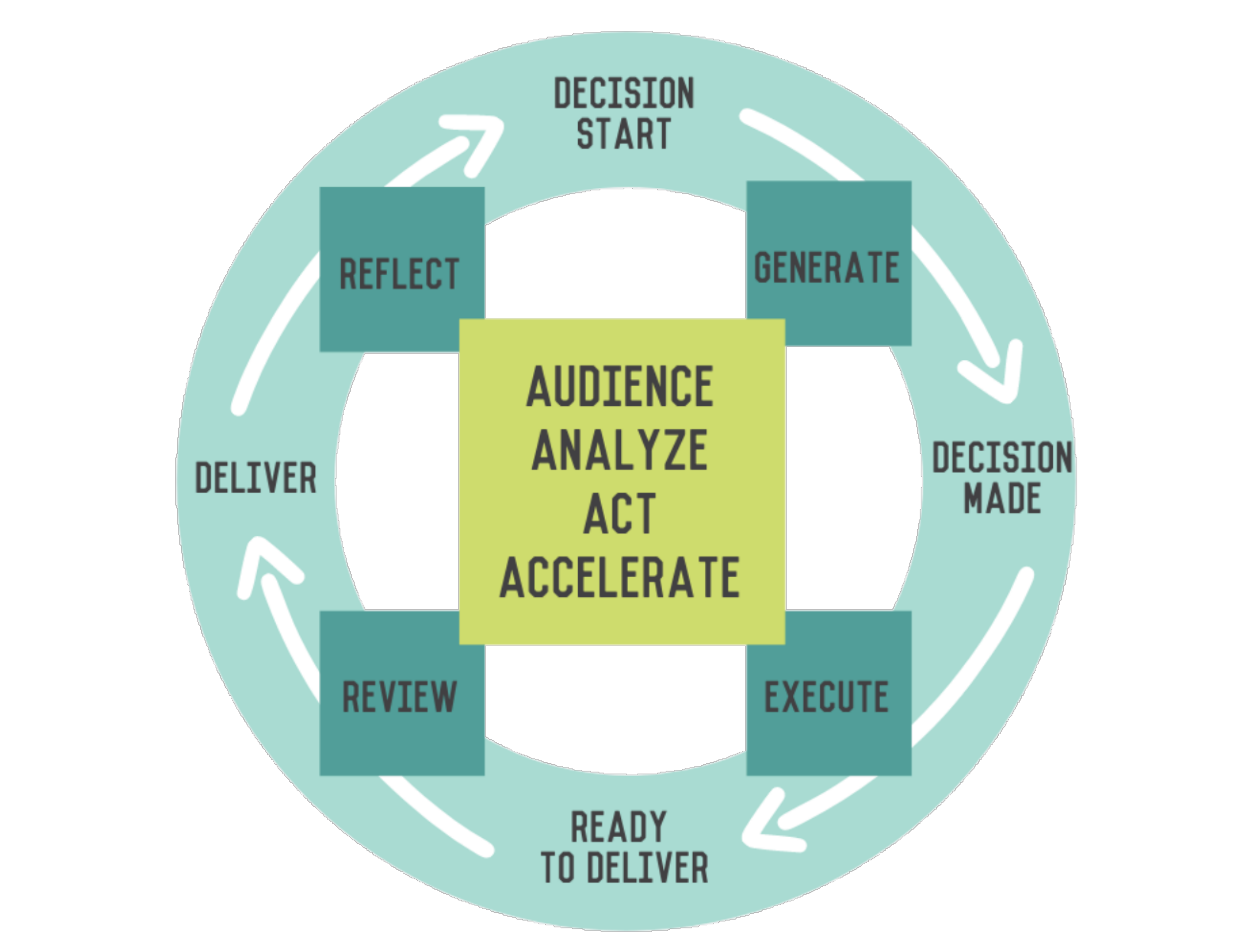}
  \caption{2D4A Framework (from \citep{2D4A})} 
  \captionsetup{justification=centering}
  \label{2D4A}
\end{figure}

This framework (see Figure \ref{2D4A}) was developed to facilitate empathy driven development. Empathy driven development helps the developers to put empathy at the centre of their work \citep{2D4A}. They have defined empathy driven development as “the practice of anchoring decisions on the people impacted by and who interact with what is produced”. The 2D4A acronym refers to Decision, Deliver, Audience, Analyse, Act, Accelerate. The 2Ds helps to understand when we should focus on implementing empathy. They are introduced with the purpose of triggering a pause to identify decision and deliverable opportunities. Decision opportunities may include what should developers do \textit{(choices)}, what could developers do \textit{(ideas)} and what could be better \textit{(improvements)}. Deliverable opportunities comprise of coding, recording information, asynchronous and synchronous communication. The 4As are categories of questions to assist developers with their work. In audience factor, the most direct person of focus, additional individuals who would be impacted, context and environment of the impacted individuals, pain considerations or the challenges faced by the impacted individual, and the gain consideration or the benefits for the impacted individual should be considered. Analyse factor refers to the aspects which should be considered before acting such as data quality, feasibility, risks/benefits, edge cases, and error mitigation. Act factor in the framework refers to the way the person should act which focuses on the medium of passing information, structure of information, discoverability of information, durability of information and artefacts which need to be created. The final factor, accelerate, refers to the reflection of an act such as learnings, impacts of sharing information, potential future collaborations and the impact of current systems (eg., processes, tools, policies) on the creation of communication artefacts. Further this process is repetitive and emphasises the significance of empathy throughout the software development process. 

\section{Techniques of Empathy} 
\label{empathyTechniquesSection}
We categorised the ways of applying empathy to improve empathy in an act as techniques of empathy. These cannot be used to integrate empathy as a process, yet can be utilised to promote empathy in the phases of a given process. There are numerous tools such as personas, EM, customer journey maps, emotional response cards, empathy interviews which belong to this category. We only describe some key techniques in this section.

\begin{figure}[h]
\centering
  \includegraphics[scale=0.2]{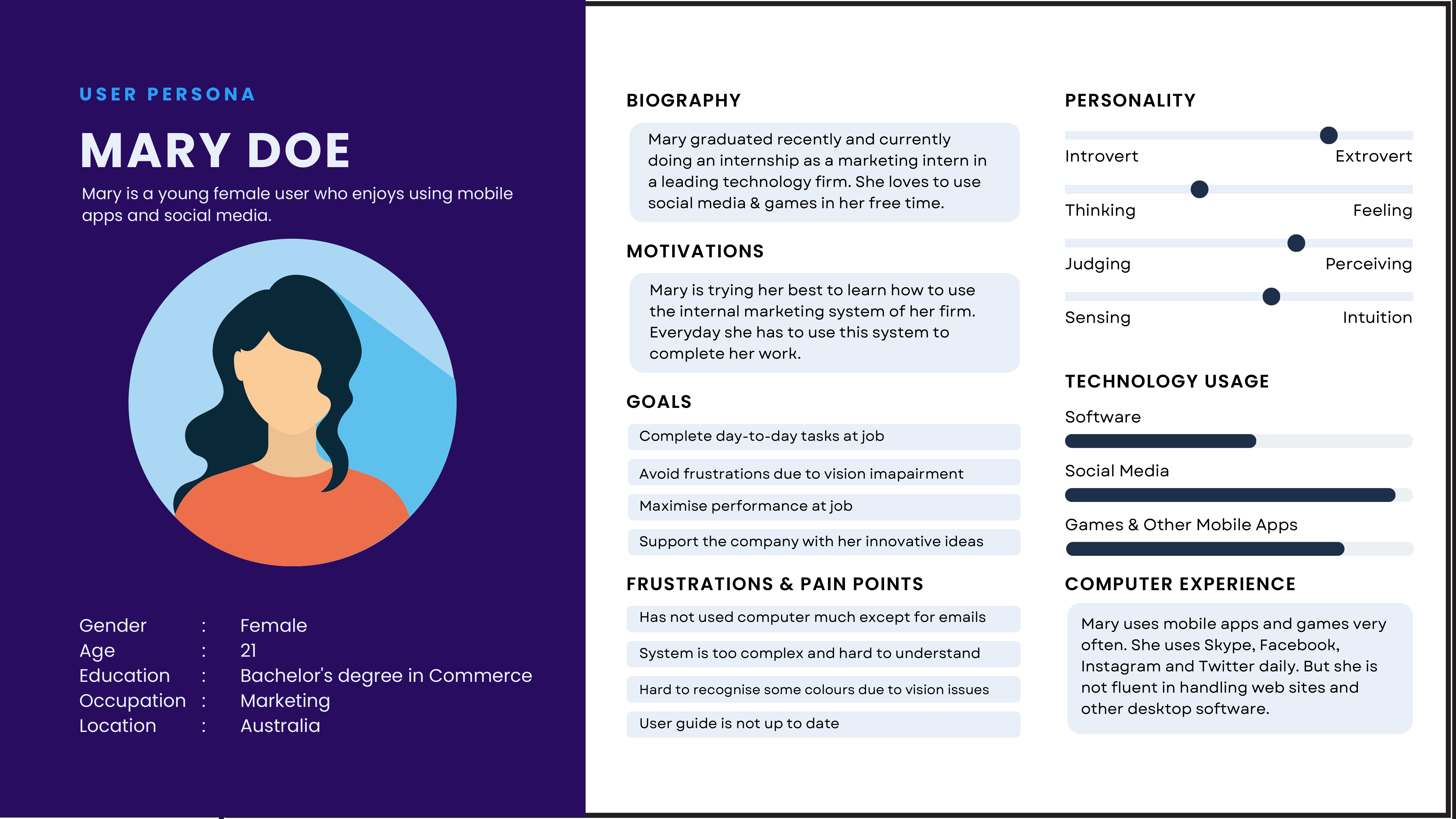}
  \caption{Sample Persona}
  \captionsetup{justification=centering}
  \label{PersonaSample}
\end{figure}

\begin{itemize}
    \item \textit{Personas}
\end{itemize}
Personas (see Figure \ref{PersonaSample}) are considered as archetypical users created based on the findings of observations or other user studies designed to explore more about the real customers and their lifestyles \citep{personas2019}. Creating personas helps to identify values, attitudes, behaviours, needs, experiences, goals, interests and even the limitations of the users. It is possible to create accurate user profiles with the use of personas which helps to better understand the target users and their world. Thus personas provide an effective way of building empathy. It is highly recommended to conduct necessary user studies before designing the personas, as the personas created using our assumptions or the imagination would not represent the real users. This is a widely used empathy tool in diverse fields such as requirements engineering, designing and marketing. 

\begin{itemize}
    \item \textit{Empathy Maps (EMs)}
\end{itemize}
EM (Fig 1, page 17, of \citep{bratsberg2012empathy}) is a tool which is heavily used to understand the user needs, which helps to develop a deeper understanding of the users \citep{em2020}. It is one of the many tools which assists in empathising by analysing the observations, and helps to identify insights about the needs of the users. There are 8 areas that are most commonly covered within an EM namely See, Say, Do, Think, Feel, Hear, Pains/Top challenges and Gains, which means EM displays what the customer is seeing, saying, doing, thinking, feeling, hearing, and what gives grief and enjoyment to the customer \citep{bratsberg2012empathy}. Hence EM helps to build an awareness on user experiences. It would be easier to identify what users see, say, do and hear but a careful consideration is needed to determine what they think, feel, their pains and gains. 

\begin{figure}[h]
\centering
  \includegraphics[scale=0.3]{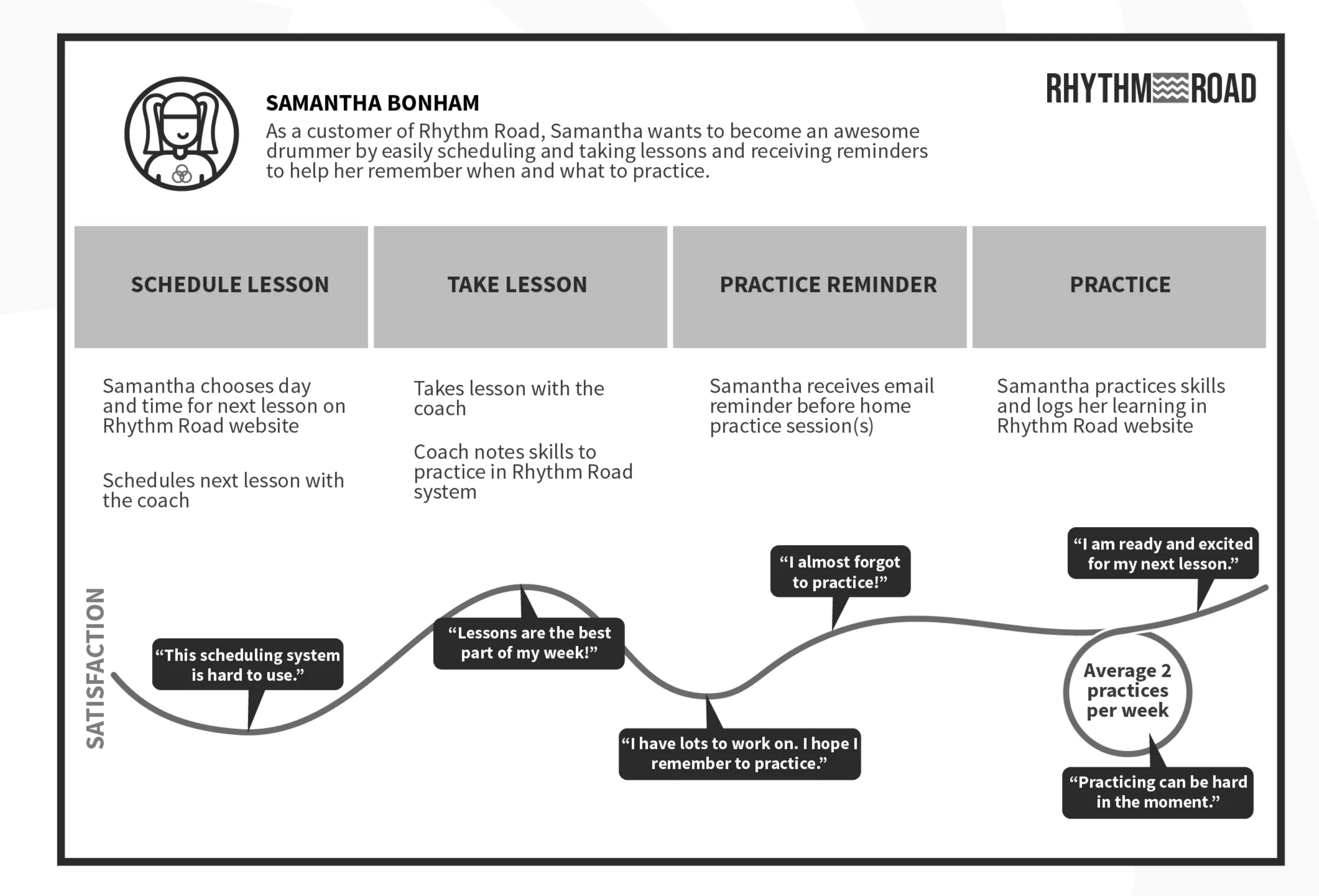}
  \caption{Sample Customer Journey Map (from \citep{komninos2018})}
  \captionsetup{justification=centering}
  \label{CustomerJourneyMaps}
\end{figure}

\newpage
\begin{itemize}
    \item \textit{Customer Journey Maps}
\end{itemize}
This is another famous tool in applying empathy which is used to understand the customer experiences over time (see Figure \ref{CustomerJourneyMaps}) \citep{komninos2018}. Customer journey map is used to plot the relationship between a customer and a product or service over time. Customer's interactions with the product or service across all media are considered when designing the customer journey map. This model helps to understand how the experience of customer develop with the time and empathise with their experience. Further it is used to analyse whether the customer experiences fulfil the customer expectations or not. It is advisable to do proper user research to obtain accurate user feedback before designing customer journey maps. If not, the product/service development organisation would not be able to get the maximum benefit out of created maps as they would not represent the real user experiences.   

\begin{figure}[h]
\centering
  \includegraphics[scale=.4]{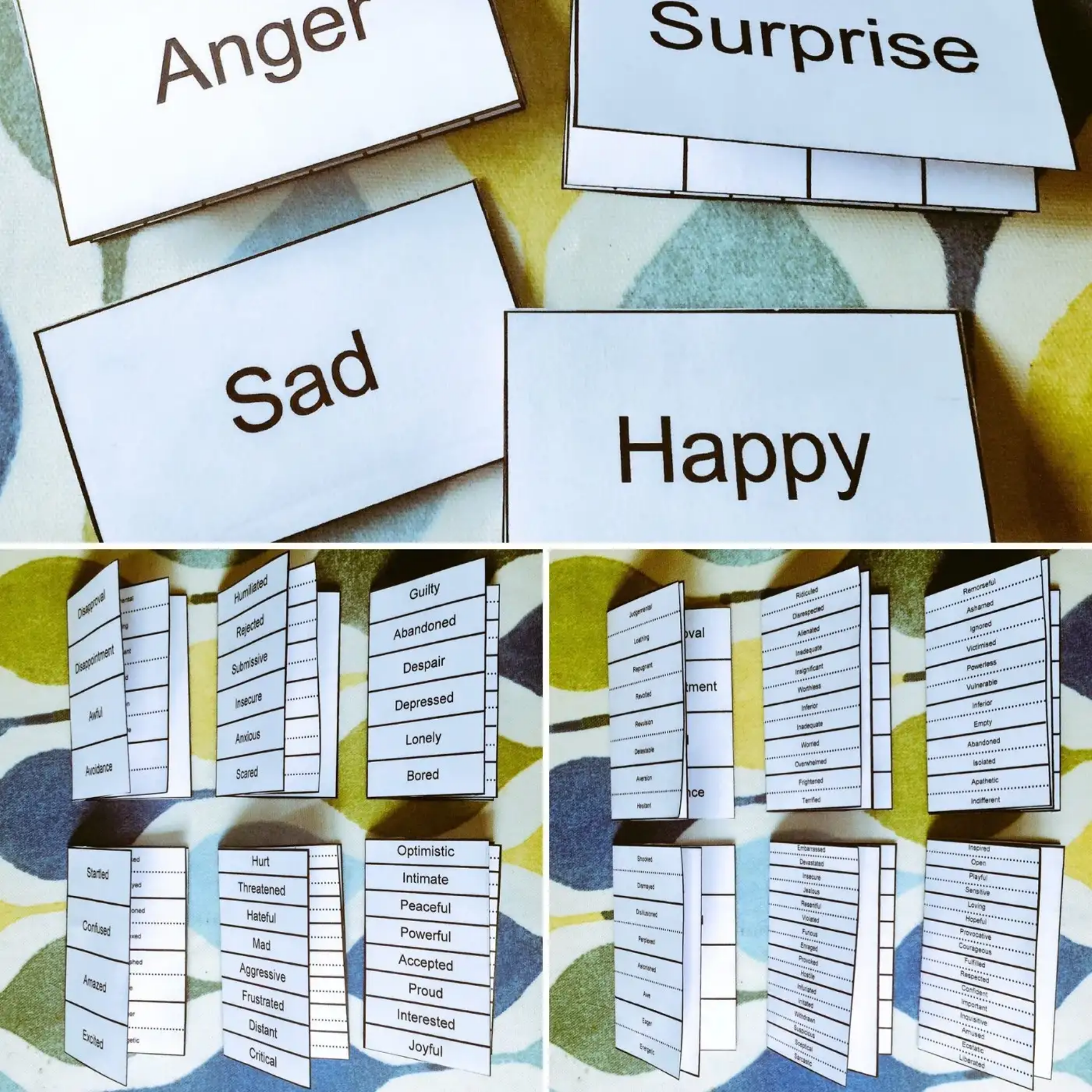}
  \caption{Emotional Response Cards (from \citep{somerville2016emotioncards})}
  \captionsetup{justification=centering}
  \label{emotionalResponseCards}
\end{figure}

{\begin{itemize}
    \item \textit{Emotional Response Cards}
\end{itemize}
Emotional response cards (see Figure \ref{emotionalResponseCards}) which are also known as emotion cards are used for demonstrating common human emotions like happy, sad, angry, surprised, anxious, confused, worried and tired. Emotional response cards are simply a set of cards with emotions which allow individuals to deeply analyse their different emotional states. This method is often used in treating patients with autism and when studying languages. Also when considering the SE domain, this is a popular technique used in UX design. This technique is used to facilitate more in depth and clear discussions of emotional issues in designs. One of the major concerns observed in design is lack of common vocabulary to discuss emotions. Emotional response cards solves this issue by allowing individuals to create a shared vocabulary to describe emotions. These cards assist people to share and develop emotional meanings at a given time/place with a person or a group. The facilitator of the session can decide where and in which occasions they intend to use these cards. Usually these cards are provided as an aid to the participants and they are instructed to choose a card that reflects their emotions regarding the given matter. Emotional response cards empower participants to discuss their emotions in a clear way which allow them to empathise with others in a better way \citep{somerville2016emotioncards, pahuja2014emotion}. Further the product reaction cards developed by Microsoft is a similar technique to emotional response cards. \citep{boehm2010organised}. They developed a set of 118 words to be used in user testing workshops in order to assist individuals in articulating their emotional responses to a product. These words include both positively and negatively worded terms like 'Trustworthy', 'Understandable', 'Frustrating' and 'Overwhelming'. The facilitator of the user testing session can instruct users to select the cards that best describes how using the the product made them feel. As the next step, facilitator can ask user to narrow down their selection to five cards and there will be an interview following this card selection process to understand the reasons for selecting those cards. This technique can be used for both existing and new products which can be used to identify improvement areas in products by capturing experiences and feelings of users \citep{boehm2010organised, Turner2016Capturing}.}

{\begin{itemize}
    \item \textit{Empathy Interviews}
\end{itemize}
Empathy interviews are one-on-one conversations used to better understand users by exploring diverse lived experiences of people. In these interviews, empathy is considered as the ability to understand user's perspective or experiences despite the conflicts with researcher's experiences. They are vital in understanding the experience of individuals who are directly impacted by the process, service or program. Usually empathy interviews contain open-ended and story-based questions which are used to obtain information about participants which assists in discovering unacknowledged needs. These interviews facilitate more in-depth discussions on the lived experiences of users compared to traditional interviews. Also empathy interviews assist in assuring that these lived experiences are considered in decision making and other actions related to that particular process, service or program. Empathy interviews are useful in discovering the human-centered improvements, determining the issues faced by individuals and identifying the root causes of issues from a community point of view \citep{nelsestuen2020empathy, lochmiller2023using}.   }

\section{Measures of Empathy} 
\label{empathyMeasuresSection}

\begin{figure}[h]
\centering
  \includegraphics[scale=.55]{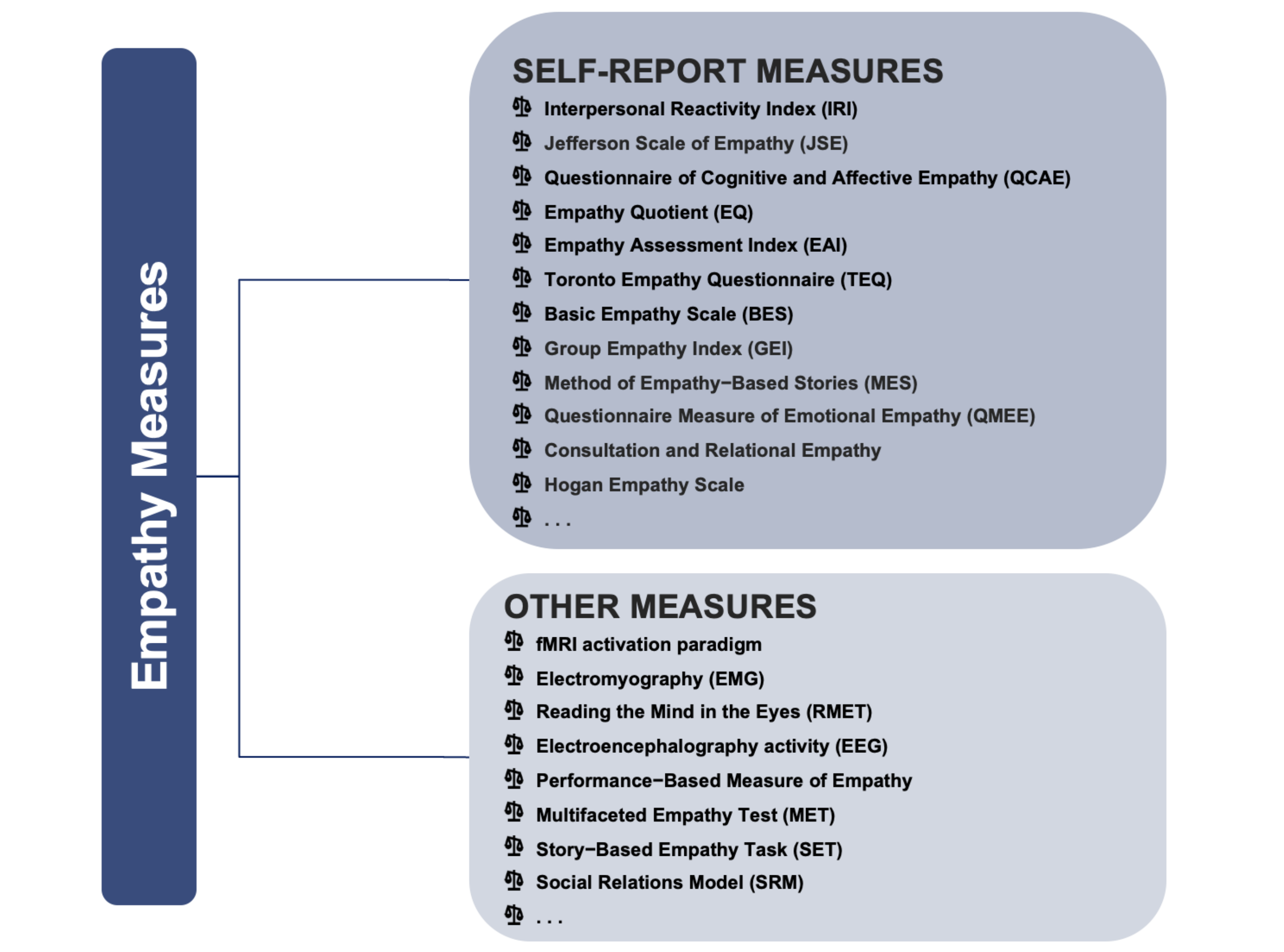}
  \caption{Measures of Empathy}
  \captionsetup{justification=centering}
  \label{empathyMeasures}
\end{figure}
Many constructs to measure empathy have been implemented over the years as shown in the Figure \ref{empathyMeasures}. Most of these constructs are self-report measures. A few are measurement techniques such as fMRI activation paradigm, Electroencephalogram (EEG), and Electromyography (EMG) which do not belong to self-report measures  \citep{ilgunaite2017measuring}. In this section we only discuss the most widely used self-report measures.

\begin{itemize}
    \item \textit{The Jefferson Scale of Empathy (JSE)}
\end{itemize}

\par JSE is a 20-item index that was developed to value empathy in the environment of medical education and the care of patients \citep{ilgunaite2017measuring}. JSE is divided into three components: perspective taking, compassionate care and "walking in patient’s shoes". {The Perspective Taking subscale measures the tendency to resonate with the views of others whereas compassionate care subscale assesses the ability understand experiences of the patients. The walking in patient’s shoes subscale focuses on the ability to think from patient's perspective without losing sight of one’s own role and responsibilities.} The questionnaire is answered by a Likert scale of 7-points from strongly disagree(1) to strongly agree (7) and the reliability of JSE has been shown to be quite high.  {JSE is the generic or the original version of this scale and later its content was modified to design different versions} as Jefferson Scale of Physician Empathy (JSPE), Jefferson Scale of Patient’s Perceptions of Physician Empathy (JSPPPE) and Jefferson Scale of Empathy − Health Profession Students version (JSE−HPS). {The JSPE assesses different components of empathy between physicians in patient−care environment and the JSPPPE measures empathy of physicians from patients' perspective. The JSE−HPS version is used to measure empathy of students in healthcare environments.} Many researchers have validated the effectiveness of JSE in different contexts \citep{ilgunaite2017measuring, ward2009reliability, shariat2013empathy, roh2010evaluation, paro2012brazilian, chatterjee2017clinical, hojat2015eleven, alcorta2016cross, jeon2015assessment, Hojat2016}. 

\begin{itemize}
    \item \textit{Interpersonal Reactivity Index (IRI)}
\end{itemize}
\par IRI is a 28-item questionnaire to measure different reactions and personal experiences of one individual while observing the other \citep{ilgunaite2017measuring}. It is designed to measure different empathic tendencies such as Perspective Taking, Fantasy, Empathic Concern and Personal Distress. {The perspective taking scale measures the tendency
to shift to the psychological point of view of others \textit{(eg: "Before criticizing somebody, I try to imagine how I would feel if I were in their place")} whereas the fantasy scale assess the tendency to imaginatively transpose oneself into fictional situations \textit{(eg: "I really get involved with the feelings of the characters in a novel.")}. The personal distress scale measures the tendency to experience fear or anxiety in response to extreme distress in others \textit{(eg: "In emergency situations, I feel apprehensive and ill-at-ease")} and the empathic concern scale assesses the tendency to feel compassion or sympathy for others who are less-fortunate \textit{(eg: "When I see someone being taken advantage of, I feel kind of protective towards them").}} Each of these tendencies is made up of seven items. The questions are answered using a 5-point Likert scale. This is the most widely used measure of individual levels of empathy in social psychology \citep{ilgunaite2017measuring, marshall2021measuring}. 

\begin{itemize}
    \item \textit{Questionnaire of Cognitive \& Affective Empathy (QCAE)}
\end{itemize}
\par QCAE is a 31-item questionnaire that was developed to measure affective and cognitive empathy, and these two components are divided into five different subscales \citep{ilgunaite2017measuring}. The cognitive empathy component consists of two subclasses, Perspective Taking and Online Simulation, comprising 10 and 9 items respectively. Perspective taking evaluates how one person is able to see the situation from another person’s perspective {\textit{(eg: "Other people tell me I am good at understanding how they are feeling and what they are thinking")}} and online simulation assesses the ability of a person to understand and mentally represent how another person is feeling {\textit{(eg: "Before criticising somebody, I try to imagine how I would feel if I was in their place")}}. There are 3 subclasses in the affective empathy component -- Emotion Contagion, Proximal Responsivity, Peripheral Responsivity. All of these subclasses comprise 4 items in each. Emotion Contagion helps to see how the person is able to reflect self−oriented emotions while noting the emotional states of others {\textit{(eg: "People I am with have a strong influence on my mood")}}. Proximal Responsivity assists in measuring people’s emotional reaction to the moods of another person, who is physically or emotionally close to them {\textit{(eg: "It affects me very much when one of my friends seems upset")}} whereas Peripheral Responsivity measures people’s emotional reaction with respect to the moods of another person, who is not close to them or a stranger to them {\textit{(eg: "I often get deeply involved with the feelings of a character in a film, play, or novel")}} \citep{reniers2011qcae}. 

\begin{itemize}
    \item \textit{Empathy Quotient (EQ)}
\end{itemize}
\par EQ is a 60−item questionnaire which was developed with the aim of measuring empathy in adults of normal intelligence \citep{baron2004empathy}. EQ has been evaluated to observe the scores of adults with high-functioning autism (HFA) or Asperger Syndrome (AS) and also to test gender differences in empathy. EQ measures three different factors including cognitive empathy, emotional reactivity and social skills \citep{ilgunaite2017measuring}. {Cognitive empathy subscale measures the ability to put oneself into the place of someone else \& adopt their perspective \textit{(eg:"I can easily work out what another person might want to talk about")} and emotional reactivity assesses the willingness to express emotions rather than to the ability to identify mental states and respond with the correct emotion \textit{(eg:"I really enjoy caring for other people")}. The social skills subscale measure the sensitivity to social situations \textit{(eg: "Friendships and relationships are just too difficult, so I tend not to bother with them")}.} Questions are answered on a scale from 0 to 2 and EQ has been shown to be very effective in measuring cognitive empathy. Apart from the empathy focused questions, EQ has 20 filler items to distract the participant from the persistent focus on empathy. This scale is designed with the purpose of making it easy to use and easy to score \citep{baron2004empathy, lawrence2004measuring}.

\section{Discussion} 
\label{discussionSection}
Most of the widely used empathy models are summarised in the previous section yet there is no mention in current state-of-the-art about an empathy model designed for the SE domain. 

\textbf{The suitability and applicability} of each of the above models for the field of SE are still not properly validated or they have not been used in SE. Our key focus being building empathy between the software developers and their users, we believe cognitive empathy is one of the major components that we should consider. We currently consider 4 common scales -- JSE, IRI, EQ and QCAE -- that consist of measures for cognitive empathy are likely to be good starting points. Even though IRI is a prominent measure of individual level empathy in social psychology \citep{marshall2021measuring}, we argue that its fantasy and personal distress components do not seem to be quite relevant for the developer-user empathy. {Fantasy scale items such as \textit{"I daydream and fantasize, with some regularity, about things that might happen to me", "I really get involved with the feelings of the characters in a novel", "Becoming extremely involved in a good book or movie is somewhat rare for me", "After seeing a play or movie, I have felt as though I were one of the characters"} and personal distress subscale items like \textit{"I sometimes feel helpless when I am in the middle of a very emotional situation", "When I see someone get hurt, I tend to remain calm", "When I see someone who badly needs help in an emergency, I go to pieces"} are some of the scale items which made us more doubtful of the applicability of IRI in a SE context.} JSE is another promising scale in the field of medicine and there are three versions of this scale which are designated to measure different instances in physician-patient relationships. Numerous research studies have been conducted to validate the usefulness of JSE and its versions in different circumstances \citep{ward2009reliability, shariat2013empathy, roh2010evaluation, paro2012brazilian, chatterjee2017clinical, hojat2015eleven, alcorta2016cross, jeon2015assessment}. {However the concentration of all the versions of JSE are towards the health professionals and patient care. For an instance JSE has scale items such as \textit{"I believe that empathy is an important therapeutic factor in medical or surgical treatment", "I try not to pay attention to my patients' emotions in history taking or in asking about their physical health", "Attentiveness to my patients' personal experiences does not influence treatment outcomes" and "My understanding of how my patients and their families feel does not influence medical or surgical treatment".} which are fully centered towards healthcare.} Due to JSE's highly focused nature towards concepts and practices in the medical domain, it is less appealing to other domains without major revisions.

EQ has been shown to be very useful to measure cognitive empathy, which makes it well-qualified to achieve our goal. However, some items in this scale -- such as "I prefer animals to humans", {\textit{"I try to keep up with the current trends and fashions", "I dream most nights", "When I was a child, I enjoyed cutting up worms to see what would happen", "It upsets me to see an animal in pain"} and \textit{"I usually stay emotionally detached when watching a film"}} -- suggests its applicability is also doubtful to our purpose without major revisions \citep{lawrence2004measuring}. QCAE is another highly accepted empathy scale which complements our need by its perspective thinking subscale items. However, some of the affective empathy related items in this scale -- such as \textit{"I am usually objective when I watch a film or play, and I don’t often get completely caught up in it"}, {\textit{"I often get deeply involved with the feelings of a character in a film, play, or novel", "I get very upset when I see someone cry" and "I usually stay emotionally detached when watching a film"}} \citep{reniers2011qcae} -- makes us question its usefulness for our problem space. 

Each of these scales have their own strengths and weaknesses which makes it challenging for researchers to choose one of them as the best-fitting scale for use in SE research contexts. Whether and how to generalise these models and what sort of studies should be conducted to evaluate the effectiveness of these models in SE contexts remains to be studied. Similarly, there are numerous untapped possibilities in evaluating the efficacy of utilising these models to enhance SE related activities such as empathy training for developers, training for Computer Science students who would be future software developers, and potentially for performance assessment in SE.  

\textbf{The term "software developer"} is used as an umbrella term in the SE context. This can vary based on the geographic location, organisation, department, project and even on the assigned job responsibilities. Depending on these factors software developers play diverse roles including but not limited to backend software development, frontend software development, requirements engineering, user experience design, quality assurance, test automation, system support and even data related services i.e., database administration, data migration. Even though our problem space includes only frontend and/or backend developers, we are debating about the potential of considering the above mentioned diverse roles played by the software developers based on the context. 

\textbf{Context of interaction} becomes a challenge when employing scales to measure empathy in SE and in other disciplines. There is an issue as to capability and adequacy of administering different empathy scales or different versions of the same empathy scale to the two target groups in our work i.e., developers and end-users. We question whether the same or different models should be employed to measure empathy based on the context of interaction, for instance for developer-developer empathy, developer-user empathy, user-user empathy. 

\textbf{Empathy training} is another interesting area to study with application to SE. Empathy training has long been practiced in medical education to build patient-clinician empathy. Research needs to ascertain whether software engineers can be trained to be more empathetic and whether there are any limitations in empathy training in an SE context. Some studies even state that the ability to practise empathy is influenced by genetics \citep{warrier2018genome}, while other researchers have argued the efficacy of empathy training \citep{teding2016efficacy}. They have found that empathy training programs are generally effective at raising empathy levels. Their findings indicated that the trainings were most effective with health professionals and university students who were compensated for their time. Along with few other future directions they have also encouraged conducting future research to find out whether empathy training can be used to improve the empathy levels of trainees except health professionals and university students, to what extent training benefits last after the training is completed, and whether compensating participants has better results than not compensating them. These studies suggest that empathy training is another profound research area as same as empathy \citep{warrier2018genome, teding2016efficacy}.

\begin{figure}[h]
\centering
  \includegraphics[scale=.58]{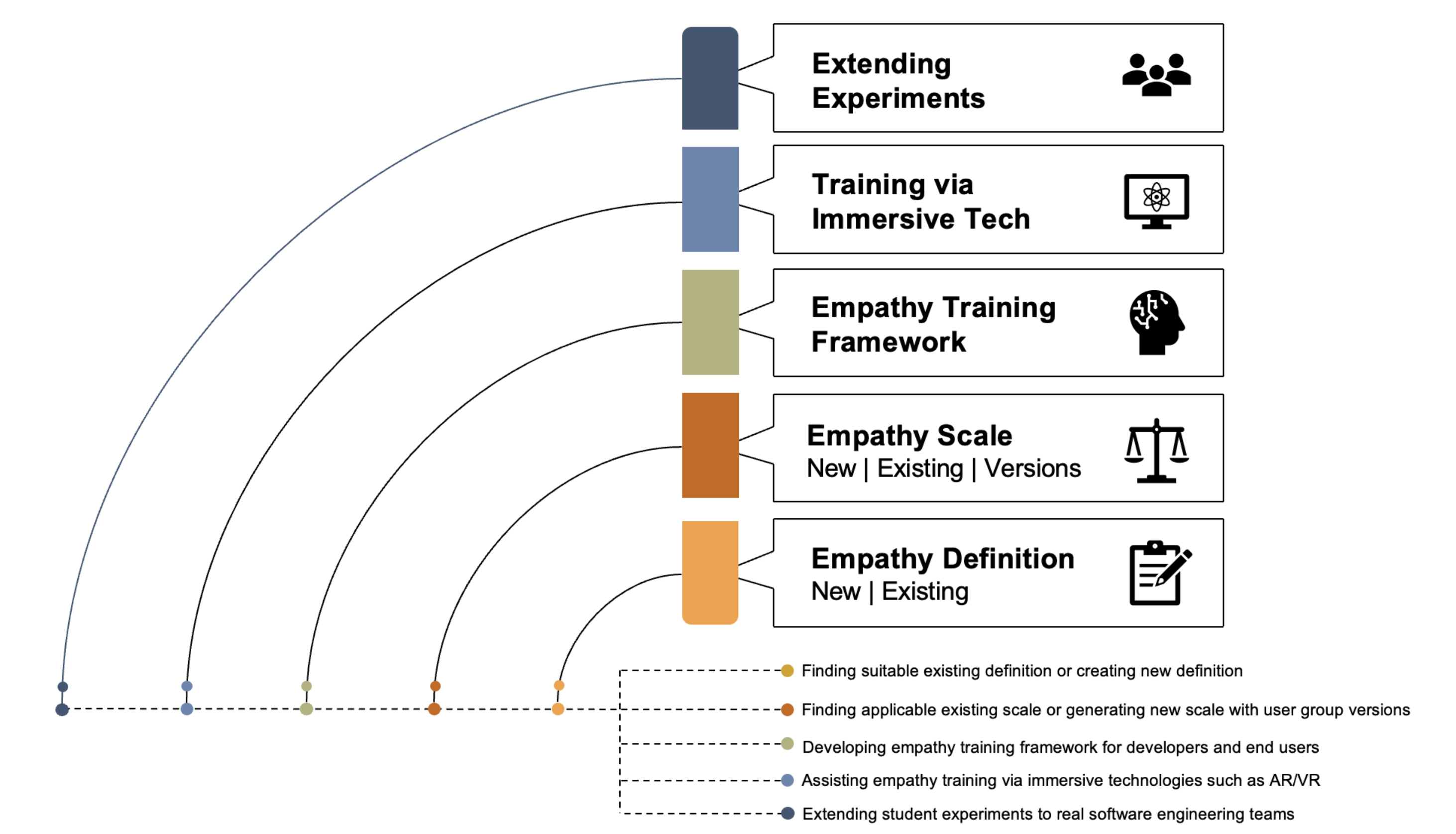}
  \caption{Research Vision on Empathy in Software Engineering Research}
  \captionsetup{justification=centering}
  \label{vision}
\end{figure}

\section{Empathy Studies in Software Engineering} 
\label{ourVisionSection}
In Figure \ref{vision} we outline some proposed studies regarding empathy in SE contexts. Our analysis suggests that empathy is still an under-researched area in SE domain. We envision that Software Engineers can benefit by incorporating empathy as another leading human aspect in SE research.
During our analysis of current state-of-art empathy research and models, we noticed that the literature lacks consensus on how empathy should be defined, measured, and applied \citep{clark2019feel}. Hence we are interested in constructing a definition for empathy which addresses the multidimensional nature of empathy in the SE context.

We are also trying to identify whether the promising empathy scales produced and used in other domains can be adopted for SE, specifically to measure empathy of software developers and end-users, and improve empathy of developers for their diverse user needs. We are exploring if developing a new empathy scale that is appealing to the field of SE is feasible, as there is some degree of uncertainty with regards to the suitability of the most promising scales in other fields as discussed above. We are also inspired to follow the JSE approach for developing separate empathy scales for developers and end-users, as we believe the same scale might not do the justice to two different user groups. 

We have a plan of developing three versions of our prospective empathy scale to administer to practitioners/developers, users and Computer Science students. We propose to further explore the development of an empathy training framework for software engineering sphere, especially targeting developers and end-users. We are also interested to explore the capability of using immersive realities such as Augmented and Virtual Reality (AR/VR) technologies to help to train people to be more empathetic towards people very different to themselves. Currently, we are working with student groups to study empathy shared between developers and end-users. In future we plan to extend this approach to the software industry where we aim to study how empathy is practised in  real software engineering teams, and how it may be further enhanced to improve software and software engineering. 

\section{Summary} 
\label{summarySection}
The purpose of developing software is to support human endeavours. It is evident that the software becomes more human-centric when human aspects are considered and included fully in all stages of the SDLC.  We present empathy as one such key human aspect which should be included in SE research and practice. Humans need to work together to produce software and developers being empathetic to their end-users and end user needs would seem to enhance software engineering outcomes. However, to date empathy has not been investigated very much in SE research. We have presented some key empathy models designed to date, predominantly used in other disciplines, and a preliminary taxonomy of empathy by incorporating most of the key models. However, there are still a lot of uncertainties in the context of empathy in SE.This includes a lack of consensus on how empathy should be defined, how the existing empathy scales should be administered, and how empathy should be integrated into SE training and SE processes. We have discussed some of the confusions we identified and also discussed our vision for future empathy research in SE. Empathy is a competitive advantage. When embedded in teams, empathy elevates the service provided by software systems and finds better solutions to the problems by getting to the heart of what is pivotal to all of us who are human i.e., connecting to each other. We hope this position paper will inspire more research into incorporating empathy into the field of software engineering.

\section{Acknowledgements}
Gunatilake, Grundy and Mueller are supported by ARC Laureate Fellowship FL190100035.

%\appendix
% \section{My Appendix}
% Appendix sections are coded under \verb+\appendix+.

% \verb+\printcredits+ command is used after appendix sections to list 
% author credit taxonomy contribution roles tagged using \verb+\credit+ 
% in frontmatter.
\printcredits

\bibliographystyle{bibliographyStyle}
% \bibliographystyle{cas-model2-names}

\bibliography{output}

% \newpage
% author1 bio
\bio{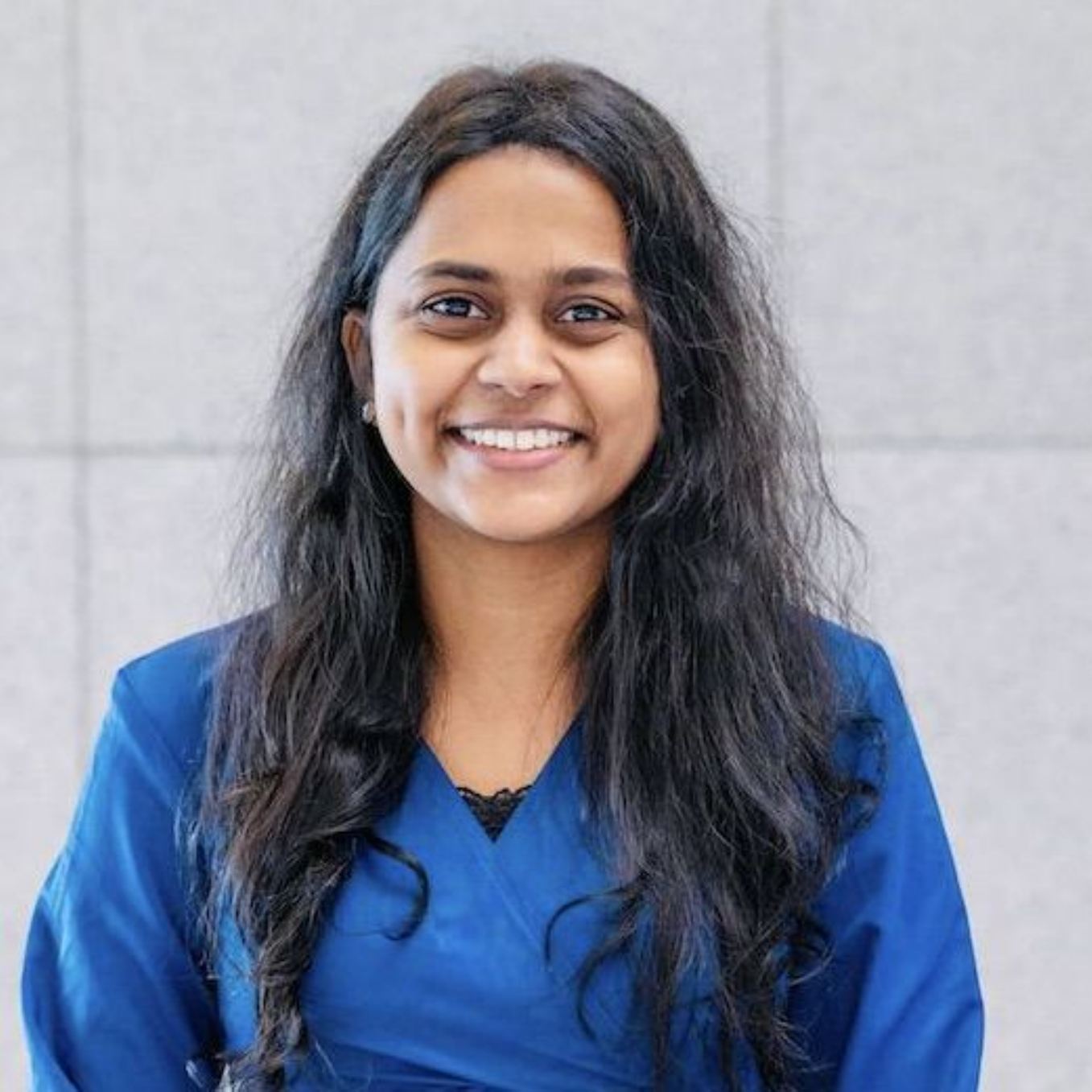}
\textbf{Hashini Gunatilake} is a PhD candidate at Monash University, Melbourne, Australia. She received her BSc (Hons.) in Information Systems degree from University of Colombo School of Computing (UCSC), Sri Lanka. Prior to her PhD candidature, she was in the software industry working as a senior quality assurance engineer. Her research interests are software engineering, human, social \& technical aspects, human-AI interaction, agile methodology, data visualisation. More details of her research can be found at, https://www.researchgate.net/profile/Hashini-Gunatilake-2. Contact her at hashini.gunatilake@monash.edu.
\endbio

% author2 bio
\bio{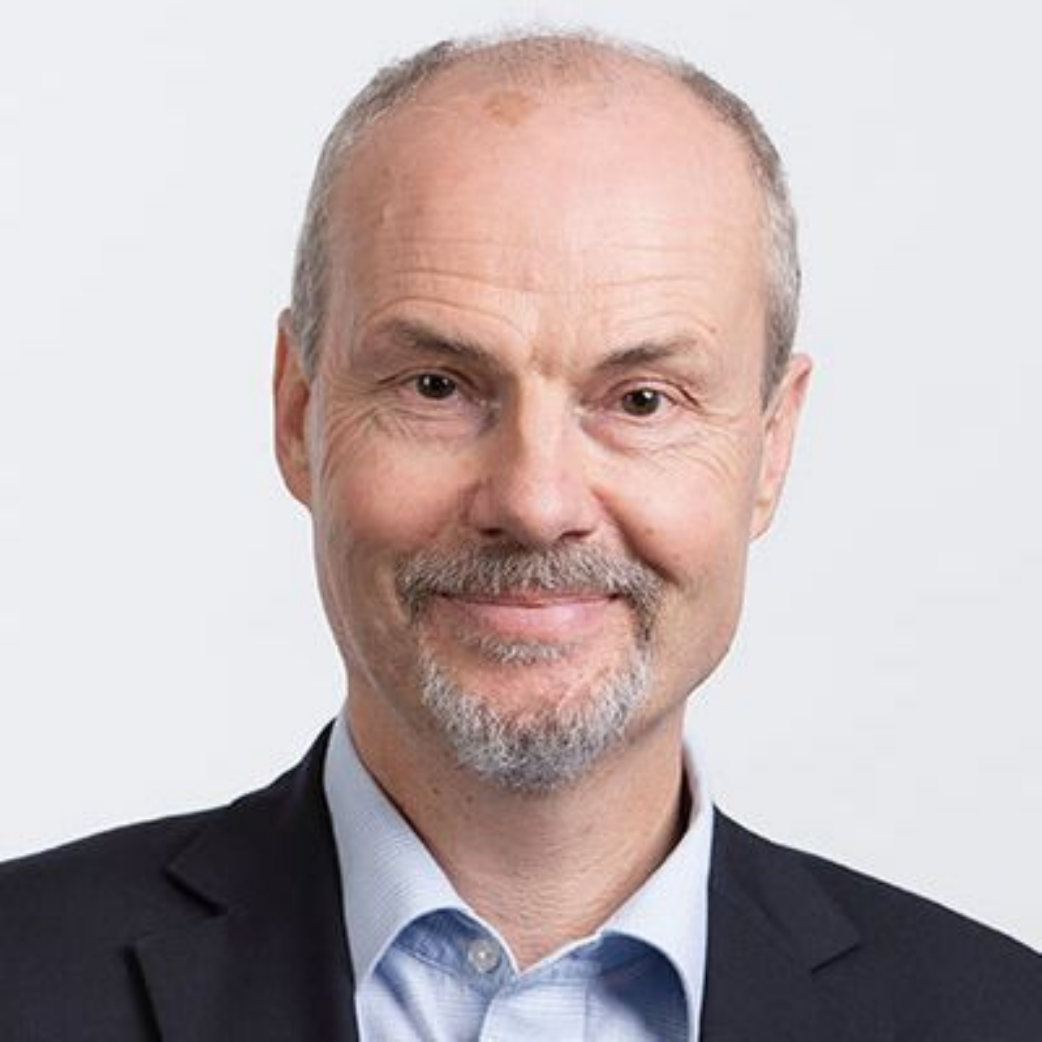}
\textbf{John Grundy} received the BSc (Hons), MSc, and PhD degrees in computer science from the University of Auckland, New Zealand. He is an Australian Laureate fellow and a professor of software engineering at Monash University, Melbourne, Australia. He is an associate editor of the IEEE Transactions on Software Engineering, the Automated Software Engineering Journal, and IEEE Software. His current interests include domain--specific visual languages, model--driven engineering, large-scale systems engineering, and software engineering education. More details about his research can be found at https://sites.google.com/site/johncgrundy/. Contact him at john.grundy@monash.edu.
\endbio
\newpage
% author3 bio
\bio{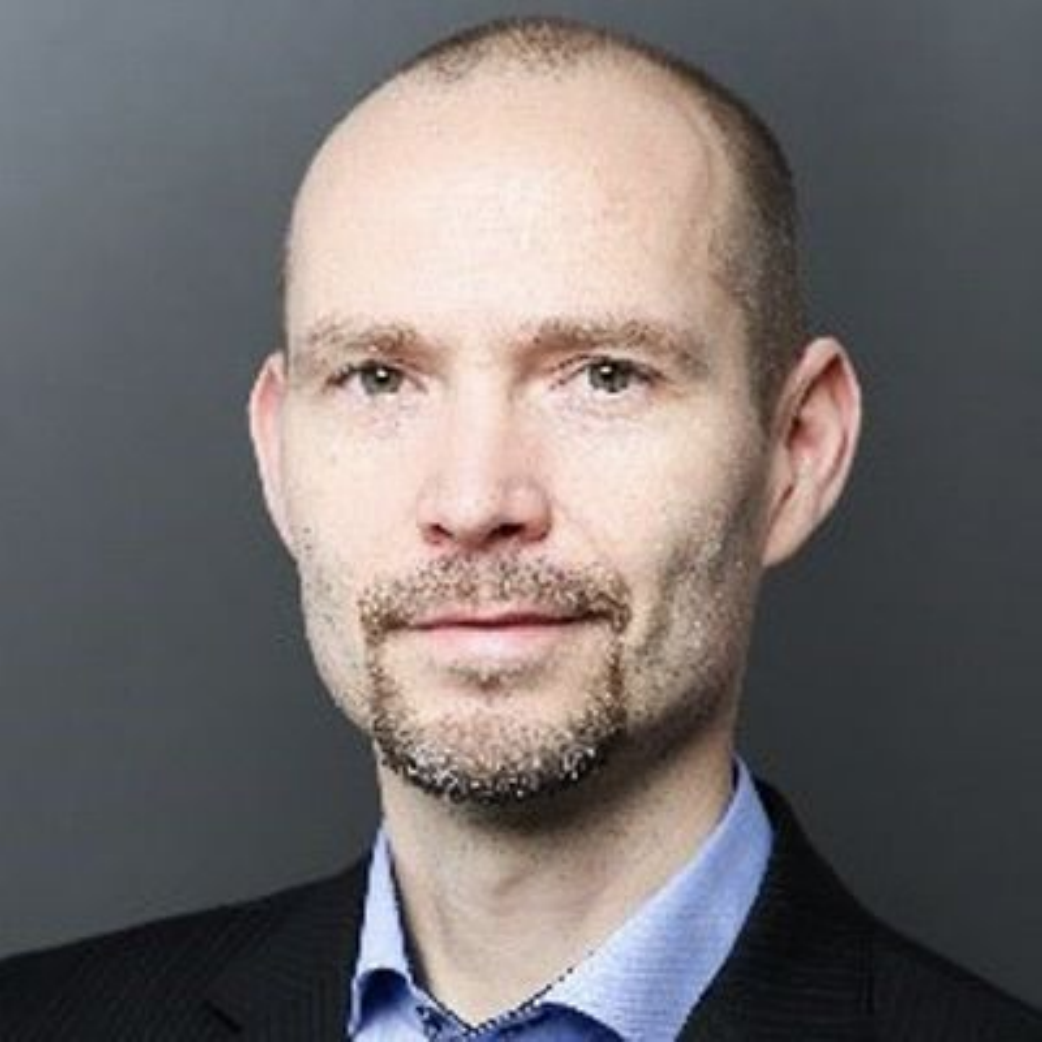}
\textbf{Ingo Mueller} is a research fellow in the HumaniSE Lab at Monash University, Melbourne, Australia. He received his PhD in computer science from Swinburne University of Technology, Australia. His research interests lie in the domain of requirements engineering including creating methods and tools that support software developers in identifying critical stakeholders, eliciting human-centric needs and capturing and modelling these needs in close collaboration with stakeholders. More details about his research can be found at https://research.monash.edu/en/persons/ingo-mueller. Contact him at ingo.mueller@monash.edu.
\endbio

% author4 bio
\bio{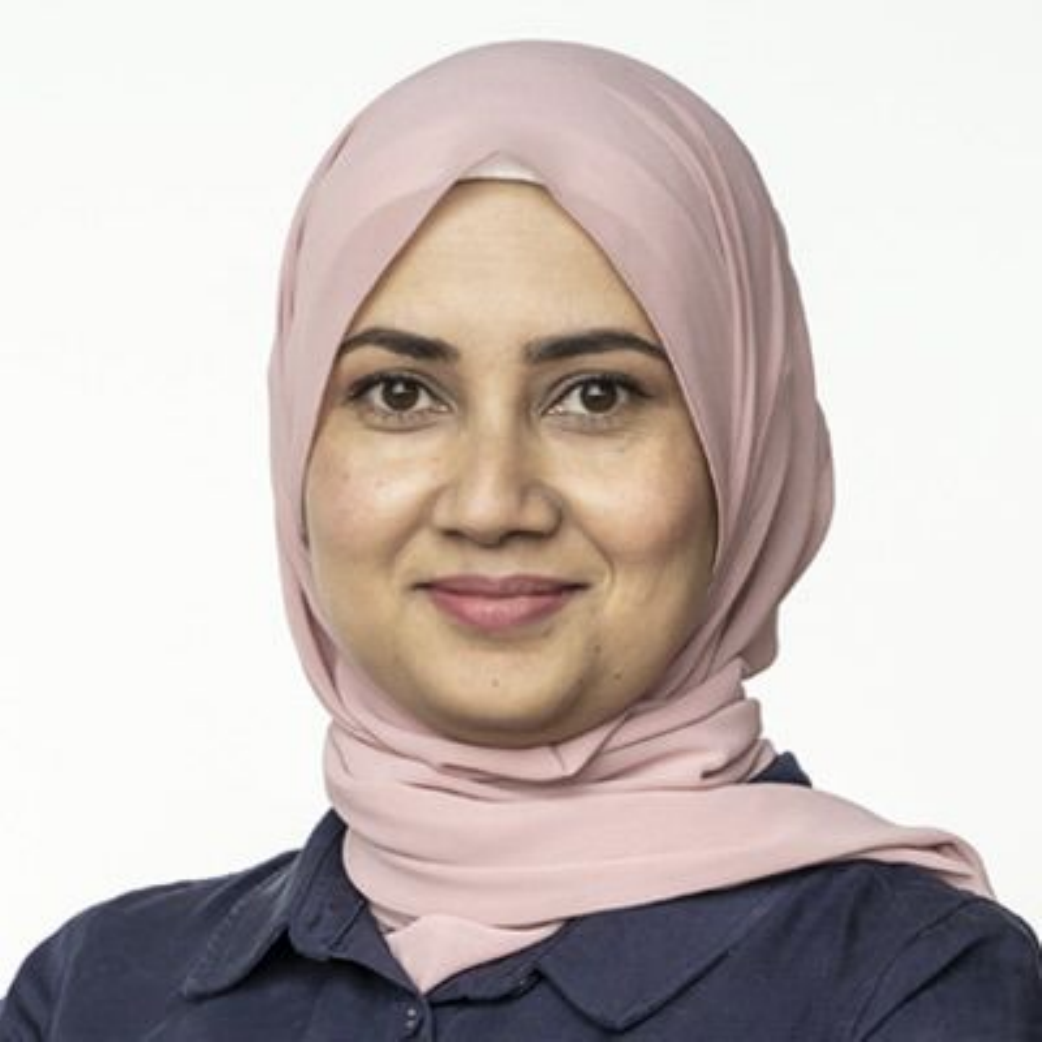}
\textbf{Rashina Hoda} is an Associate Professor in Software Engineering at Monash University, Australia. Rashina specialises in human and socio-technical aspects of software engineering and has introduced socio-technical grounded theory (STGT) for software engineering. She received an ACM SIGSOFT Distinguished Paper Award (ICSE 2017) and Distinguished Reviewer Award (ICSE 2020). She serves as an Associate Editor of the IEEE Transactions on Software Engineering and the PC co--chair of the SEIS track of ICSE2023. Previously, she served on the IEEE Software Advisory Board, as Associate Editor of Journal of Systems and Software, CHASE 2021 PC co--chair and XP2020 PC co--chair. More details on https://rashina.com. Contact her at rashina.hoda@monash.edu.

\endbio

\end{document}